\documentclass[nofootinbib,aps,prl,twocolumn]{revtex4-1}

\usepackage[utf8]{inputenc}

\usepackage{amssymb,amsmath,amsbsy,mathtools}
\usepackage{braket}
\usepackage{graphicx}
\usepackage{xcolor}
\usepackage{braket}
\usepackage{multirow}
\usepackage{dcolumn}

\usepackage{hyperref}

\usepackage{natbib}

\newcommand{\mM}{\mathcal{M}}

\newcommand{\lb}{\left(}

\newcommand{\rb}{\right)}

\newcommand{\p}{\partial}

\newcommand{\nn}{\nonumber}
\newcommand{\ee}{\end{equation}}

\newcommand{\tr}{\text{tr}}
\usepackage{graphicx}

\usepackage{graphicx}
\usepackage{dcolumn}
\usepackage{bm}

\begin{document}

\preprint{APS/123-QED}

\title{Exact Renormalization of Wave Functionals yields Continuous MERA}
\author{Samuel Goldman$^a$, Nima Lashkari$^b$, Robert G. Leigh$^a$, Mudassir Moosa$^b$}
\affiliation{$^a$ Department of Physics, University of Illinois,
1110 W. Green St., Urbana IL 61801-3080, U.S.A.}

\affiliation{$^b$ Department of Physics and Astronomy, Purdue University, West Lafayette, IN 47907, USA}

\begin{abstract}
The exact renormalization group (ERG) is a powerful tool for understanding the formal properties of field theories. By adapting generalized ERG schemes to the flow of wavefunctionals, we obtain a large class of continuous unitary networks, a special case of which includes a class of Gaussian continuous Multi-scale Renormalization Ansatzes (cMERAs). The novel feature of these generalized wavefunctional ERG schemes is allowing for modifications of the dispersion relation, which drastically changes the entanglement structure of the ultraviolet states. Through our construction, we demonstrate that cMERA can be derived from a more fundamental ``microscopic'' principle, which amounts to the usual RG principle of path integral independence, suitably adapted to quantum states of the field theory. The establishment of such a principle may provide a path forward for exploring cMERA beyond the free field regime, and for understanding the nature of entanglement renormalization intrinsically in the continuum. 

\end{abstract}

\maketitle

\section{Introduction} 
In the study of entanglement and its properties in many body systems and field theories, a well established connection has emerged between the study of tensor networks and holographic quantum gravity. The central element of this dialogue is entanglement renormalization, a powerful tool for understanding the structure of real space entanglement in lattice theories and other discrete quantum mechanical systems with spatial structure \cite{vidal2007entanglement,evenbly2015tensor}.

A particularly important example, known as the multi-scale renormalization ansatz (MERA), is a tensor network model in which local or quasi-local unitary operations act to remove entanglement between nearby sites. The lattice is then dilated by an isometry and the process is repeated, resulting in a final lattice state with similar long range correlations to the initial state, but with much simpler small scale structure. It is also of theoretical interest to consider a ``reversed'' process where entanglement is added, rather than removed.  One starts with a spatially unentangled state and uses local unitaries to entangle neighboring sites. This describes a process by which one efficiently \textit{generates} states with entanglement at many different scales, such as in the ground state of interacting QFTs. 

The connection between MERA and holography comes from the identification of the depth, or scale, of the MERA network with the radial coordinate of an asymptotically Anti-de Sitter (AdS) spacetime \cite{evenbly2011tensor,swingle2012entanglement}. Since its inception, the proposed connection between tensor networks and geometry have yielded many valuable insights into features of quantum gravity, such as the RT formula \cite{nozaki2012holographic}. 

Despite the success of tensor network models in studying holography, the program is ultimately a toy model. Because of its discrete nature, the ``dual geometry'' of MERA describes--at best--a discretized approximation of spacetime. This limitation has motivated definitions of MERA-like networks directly in the continuum. The program, designated continuous MERA (cMERA), consists of a one-parameter family of unitary maps $\hat{U}(z)$ with $z\in (1,\infty)$ and $\hat{U}(1) = \hat{I}$ \cite{haegeman2013entanglement}. Explicitly, the unitary operator takes the form of a path-ordered exponential,
\begin{equation}\label{unitary}
    \hat{U}(z) = \mathcal{P}_z\left[e^{i\lb \int_1^z \frac{dz'}{z'}\left(\hat{\mathbf{K}}(z') +  \hat{\mathbf{L}}\right)\rb}\right].
\end{equation}
    The Hermitian generator $\hat{\mathbf{L}}$ is the Weyl scaling in $d$ spacetime dimensions\footnote{We use the Weyl scaling for its simplicity, but note that our choice is different from the relativistic and non-relativistic scaling operators defined in  \cite{nozaki2012holographic}.},
\begin{eqnarray}\label{condL}
    \hat{\mathbf{L}}=\frac{d-2}{4}\int_{\vec{x}}\Big(\hat{\varphi}(\vec{x})\hat{\pi}(\vec{x}) +\hat{\pi}(x) \hat{\varphi}(\vec{x})\Big),
\end{eqnarray}
and the operator $\hat{\mathbf{K}}(z)$ implements the analog of an entangling operation on neighboring sites:
\begin{equation}\label{cond3}
    \hat{\mathbf{K}}(z) = \frac{1}{4}\int_{\vec{x},\vec{r}}  \;g_z(\vec{r}) \Big(\hat{\varphi}(\vec{x})\hat{\pi}(\vec{x}+\vec{r}) +\hat{\pi}(\vec x+\vec{r}) \hat{\varphi}(\vec{x})\Big),
\end{equation}
with $g_z(\vec{r})$ some quasi-local smearing function that falls off at large $r$. The entangling kernel $g_z(\vec{r})$ is chosen such that if $\ket{\Lambda}$ is a field theory state with no real space entanglement, then 
\begin{equation}\label{cond1}
     \hat{U}(z) \ket{\Lambda} \stackrel{z\to\infty} {\longrightarrow}\ket{\Omega},
\end{equation}
where $\ket{\Omega}$ is some highly entangled state such as the ground state of a relativistic field theory. In this way, cMERA mimics the state preparation of lattice MERA networks. 

The forms of MERA and cMERA are motivated by qualitative features of the Wilsonian RG. However, unlike MERA, which is constructed from an underlying microscopic principle, the definition of cMERA networks is essentially \textit{ad hoc}: it is built to connect two particular states along a flow in a way which is reminiscent of the end points of MERA. Moreover, unlike its discrete partner, which has seen success in describing ground states of strongly coupled lattice theories, cMERA has generally resisted attempts to go beyond Gaussian states and free field theories. These two problems are not entirely unrelated: without a well defined microscopic principle, it is not manifest what the appropriate steps are to include interactions in a cMERA.

The question then, of whether one can define cMERA from an underlying ``RG principle'' directly in the continuum, is one of great interest for developing a complete picture of continuous entanglement renormalization, and ultimately for applications of cMERA in holography and gravity. In this paper, we will argue that such a principle indeed exists, and it is in fact precisely the usual notion of an RG principle used in the Exact Renormalization Group (ERG) \cite{polchinski1984renormalization}, suitably adapted to field theory states. The groundwork for our analysis was laid in \cite{fliss2017unitary,leigh2015exact,leigh2014holographic}. By modifying the renormalization scheme of Polchinski, it was shown that one can derive an ERG of wavefunctionals whose flow is unitary and has precisely the form of eq. (\ref{unitary}). However, this flow lacked some important properties of cMERA: the starting point of the unitary flow did not yield an unentangled state as in (\ref{cond1}); and the disentangler was completely non-local in position space, unlike the desired quasi-local smearing function of (\ref{cond3}).

To overcome the challenges of constructing cMERA from the ERG, it is useful to note that in many cMERA models, the UV physics is modified in such a way that it is effectively non-relativistic. 
In particular, the ground state $\ket{\Omega_{\textrm{nr}}}$ of a non-relativistic boson takes the form
\begin{equation}\label{unentangle}
    \langle \varphi \vert\Omega_{\textrm{nr}}\rangle  \sim \exp\left(-\#\int_{\vec{x}} \varphi(\vec{x}\,) \varphi(\vec{x}\,)\right) .
\end{equation}
As we can see from the right hand side, this wavefunctional factorizes in position space, and thus has no real space entanglement. This motivates us to generalize the ERG by considering smooth cut-off functions that modify the dispersion relation, allowing for the UV to be non-relativistic.

In this work, we implement ERG schemes that are general enough to include, as special cases, cMERAs, i.e.,  the unitary flows we derive have the necessary conditions of (\ref{unitary}), (\ref{cond1}), and (\ref{cond3}) with quasi-local kernels $g_z(\vec{r})$. The paper is structured as follows. In Section II, we review the adaptation of the Polchinski ERG scheme to wavefunctionals. In Section III, we introduce the modified regulating scheme which implements cMERA, and compute the resulting RG flow equations from the path integral. Appendix \ref{app:canonical} includes a discussion of the generalized ERG in the Hamiltonian picture that makes the connection to cMERAs more manifest. In Section IV, we conclude and comment on generalizations and applications.

\section{Review of the ERG for States}

We begin with a brief review of the adaptation of ERG to the flow of states in the singlet sector of the $O(N)$ model as constructed in \cite{fliss2017unitary}. The main principle is to implement an ERG regulator in ``sliced'' path integrals that correspond to the wave functionals of quantum states. In local field theories, this path integral integrates over field configurations on half of spacetime ${\cal M}_-$ in the past of a spacelike surface $\Sigma$ (which we take for simplicity to be a constant-time hypersurface). Arbitrary quantum states are obtained by operator insertions in Euclidean time, and a basis is chosen by fixing arbitrary field configurations $\varphi(\vec x)$ on $\Sigma$. 

Since in this paper we are interested in establishing a precise connection between the ERG and existing Gaussian cMERA constructions, we will confine our attention to the (regulated) free fixed point of the $O(N)$ scalar model in $d$ spacetime dimensions. A useful trick that was introduced in \cite{fliss2017unitary} is to write the generator of quantum states that are created by arbitrary singlet operators acting on the ground state. 
The corresponding path integral can be written 
\begin{eqnarray}\label{pathintegralexcited} &&\braket{\varphi(\vec x)|\Omega[b]}=\int [\mathcal{D}\phi]_{\phi(-i\infty,\vec{x})\to 0}^{\phi(0,\vec{x})=\varphi(\vec{x})}\:\: e^{i (S_0+S_I+S_b+\mathcal{U}_b)}
\end{eqnarray}
where\footnote{We will often use abbreviated DeWitt notation 
$(A\cdot \phi)(x):=\int_y A(x,y)\phi(y)$, and will leave implicit the contracted $O(N)$ indices.}
\begin{equation}
S_b=\int_{\cal{M}_-} \phi\cdot  b\cdot \phi
\end{equation}
and we have implemented a Dirichlet boundary condition on $\Sigma$ to select the corresponding basis for the Hilbert space. If we fix the distribution $b(x,y)$ to some value in Euclidean time, then the resulting state is Gaussian; importantly, however, note that more generally $b(x,y)$ can be regarded as a source, and so one may construct arbitrary non-Gaussian states by taking derivatives of $\braket{\varphi(\vec x)|\Omega[b]}$. Using this formalism, the ERG is interpreted as a flow in the space of singlet states, coordinatized by $b(x,y)$. 

In \eqref{pathintegralexcited}, the integration is over arbitrary field configurations in the interior of ${\cal M}_-$ while $S_0$ is a suitably regulated kinetic term that we will describe below. ${\cal U}_b$ is a field-independent term that sets the normalization of the path integral. By $S_I$ we denote interaction terms which, in the context of the $O(N)$ model, should be taken to include {\it all} multi-trace singlet operators. Indeed, with regard to the generating functional, one finds that the exact ERG equations close only on the full set of such operator sources. An important exception is the choice $S_I = 0$, which constitutes a consistent truncation of the ERG flow equations. As we have noted, such a truncation is sufficient for describing existing cMERA constructions, and we take this choice for simplicity. However, there is no obstruction, in principle, to including a non-zero $S_I$; we will comment further on the effect of multi-trace operators and interactions in the discussion section.

In contrast to the partition function that is invariant under the RG flow, the wave functionals evolve as a function of scale. An ERG regulated action can be written in the form\footnote{We recall from \cite{fliss2017unitary} that the term with support on $\Sigma$ is required in order for the boundary conditions to be consistent. Indeed, classically, the variation of the action has the form $\delta S_0[\phi]=-\int_\Sigma\,\delta\phi\cdot K^{-1}\cdot\p_t\phi$ on shell, and so Dirichlet conditions on $\Sigma$ are consistent. Allowing time derivatives in the regulator function would spoil the canonical structure.}
\begin{equation}\label{actionreg}
    S_0[\phi,\varphi] = -\frac{1}{2z^{d-2}}\left(\int_{\mM_-} \phi\cdot  D^2\cdot \phi-\int_{\Sigma} \varphi\cdot  K^{-1}\p_t\cdot \phi\right)
\end{equation}
where $K$ is a regulator and the differential operator $D^2$  was taken in  \cite{fliss2017unitary} to be of the form
\begin{eqnarray}\label{regulatepropagator}
    &&D^2:=K^{-1}(z^2\vec{\nabla}^2/M^2)(-\p_t^2+\vec{\nabla}^2)\nn\\
     &&\lim_{s\to \infty} K(s)\to 0,\qquad \lim_{s\to 0} K(s)=1\ .
\end{eqnarray}
Here, we regard $M$ as an ultraviolet cutoff scale on momenta, and $z$ has been introduced so that $M/z$ plays the role of a renormalization scale.
Importantly, we have chosen $K$ such that it regulates large spatial momentum $\vec{p}^2$, as opposed to the relativistic invariant $p_\mu p^\mu$. With this choice $S_0$ has no more than two time derivatives in the bulk and one time derivative on $\Sigma$. This is consistent with a proper symplectic structure in phase space.

The regulated action in (\ref{actionreg}) is symmetric under the transformation
\begin{eqnarray}\label{symmetry2}
    z\to \lambda z,\qquad M\to \lambda M,\qquad \phi\to \lambda^{(d-2)/2}\phi\ .
\end{eqnarray}
This is a full quantum symmetry of the wave functional 
\begin{eqnarray}\label{symmetry}
    \braket{\varphi|\Omega[b, M, z]}=  \braket{\lambda^{(d-2)/2}\varphi|\Omega[b, \lambda M , \lambda z]} ,
\end{eqnarray}
if the Jacobian coming from the path integral measure is absorbed into $\mathcal{U}_b$. 
This generalizes to a much larger background symmetry involving a bi-local linear action on the fields, but we will suppress some of these details here.  We regard \eqref{symmetry} as a background Ward identity.

There are two steps involved in the ERG. For the details of the calculations involved, the reader may consult \cite{fliss2017unitary} and appendix \ref{app:varyM}.
The first step is to lower the cutoff $M\to M/\lambda$. After some formal manipulations, one finds that the infinitesimal transformation can be written as a differential operator acting on the wave functional,
\begin{equation}\label{firststep}
-M\p_M\braket{\varphi|\Omega[b]}=\bra{\varphi} \lb\beta [b]\frac{\delta}{\delta b}+i\hat{\mathbf{K}}\rb \ket{\Omega[b]}
\end{equation}
where $\hat{\mathbf{K}}$ is a self-adjoint disentangler as in (\ref{cond3}) with
\begin{equation}\label{gdelta}
g_z=M\p_M \log K
\end{equation}
 and 
\begin{equation}
  \beta[b]=b\cdot \Delta_b\cdot b,
\qquad \Delta_b=\frac{1}{z}M\p_M (D^{-2})
\end{equation}
can be regarded as a $\beta$-function, a functional of the state source $b$.
In the second step of ERG, we make an infinitesimal symmetry transformation of the form (\ref{symmetry}) to bring $M$ back to its original value. Infinitesimally, the result can be written as
\begin{equation}\label{secondstep}
    \bra{\varphi}\lb z\p_z+M\p_M-i \hat{\mathbf{L}}\rb\cdot\ket{\Omega[b]}=0
\end{equation}
where $\hat{\mathbf{L}}$ is the scaling operator in (\ref{condL}).
Combining \eqref{secondstep} with \eqref{firststep}, we arrive at the flow equation
\begin{eqnarray}
    z\p_z\braket{\varphi|\Omega[b]}= \bra{\varphi}\lb\beta[b]\frac{\delta}{\delta b}+i\hat{\mathbf{K}}+i\hat{\mathbf{L}}\rb\ket{\Omega[b]} .
\end{eqnarray}
This equation can be integrated by enforcing the RG flow equation $z\partial_z b(z) =  \beta[b(z)]$ for the source. We say that such states have ``jumped onto the flow,'' and we obtain the unitary flow equation
\begin{eqnarray}\label{ERGflow}
  \ket{\Omega(z;b(z))}=\mathcal{P}_z\left[e^{i\int_1^z \frac{dz'}{z'}\left(\hat{\mathbf{K}}(z') +  \hat{\mathbf{L}}\right)}\right]\ket{\Omega(1;b(0))}.
\end{eqnarray}
As was remarked in \cite{fliss2017unitary} and recapitulated in the introduction, the ERG flow in (\ref{ERGflow}) resembles the unitary flow in the cMERA construction, but differs from the cMERA flow in some crucial aspects. In the next section we will modify the ERG construction to accommodate the additional features of cMERA.

\section{Non-relativistic flows}\label{sec:nonrel}
As we noted in the introduction, the UV limit of many cMERA models resembles the theory of a non-relativistic boson. Although the regulated action in (\ref{actionreg}) is not strictly Lorentz invariant due to the regulating function, it retains the dispersion relation of a Lorentz invariant theory. A crucial feature of the ERG is the freedom to choose the regulating function. In particular, there is no \textit{a priori} reason to pick a regulator which respects such a dispersion. Thus, the modification that we make in this paper is to generalize the regulator so that 
\begin{equation}\label{actionregnew}
    S^{\textrm{nr}}_0[\phi,\varphi] = -\frac{1}{2z^{d-2}}\int_{\mM_-} \phi\cdot \tilde D^2\cdot \phi-\frac{1}{2z^{d-2}}\int_{\Sigma} \varphi\cdot  K_0^{-1}\p_t\cdot \phi
\end{equation}
with
\begin{eqnarray}
    \tilde{D}^2 &=&-K_0^{-1}\partial_t^2+ K_s^{-1}\vec{\nabla}^2\nn\\
     &=& K_s^{-1}(-v_{ph}^{-2}\p_t^2+\vec{\nabla}^2).
\end{eqnarray}
Here, we have introduced two distinct functions which we call $K_0$ and $K_s$ that regulate the temporal and spatial parts of the free action, respectively. Now the regulated action has a dispersion controlled by the ratio of the two regulating functions:
\begin{eqnarray}\label{relate}
    \omega=v_{ph}\: p,\qquad v_{ph}=\sqrt{K_0/K_s}\ .
\end{eqnarray}
$K_s$ plays the role of a regulator for momentum modes of the field, and we take it to be a function of $zp/M$ with $K_s(u)\to 0$ at large $u$ and $K_s(u)\to 1$ at small $u$.

\begin{figure}[t]
\centering
     \includegraphics[width=.8\linewidth]{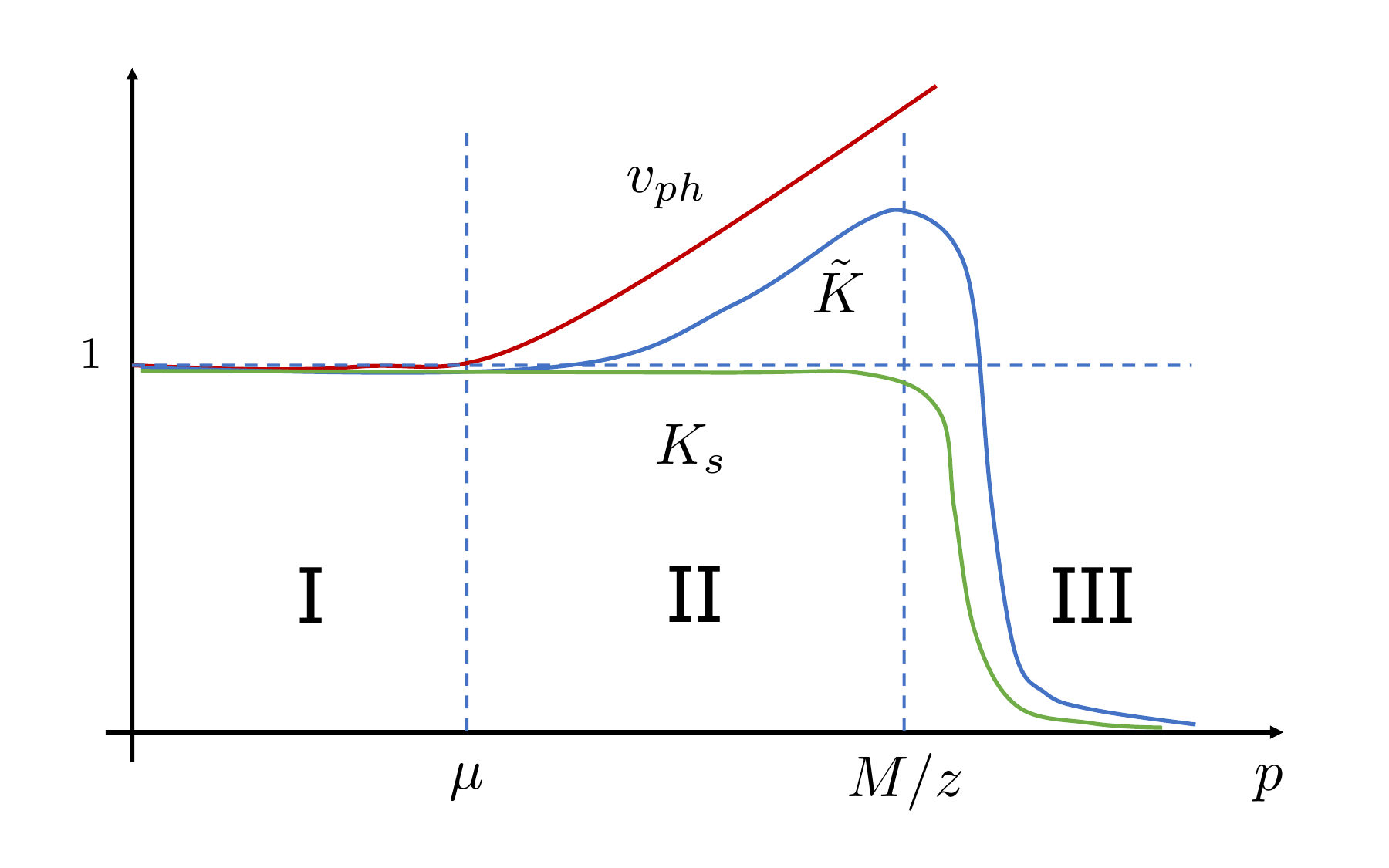}
\caption{The smooth cut-off function $K_s$ suppresses the high momentum modes above $p>M/z$. The phase velocity $v_{ph}$ transitions from relativistic $v_{ph}=1$ to the non-relativistic $v_{ph}\sim p$ at some cross-over scale $p\sim \mu$. In non-relativistic ERG, the combination $\tilde{K}=K_sv_{ph}$ controls the locality of the disentangler which can be smoother than $K_s$.}
\label{fig1}
\end{figure}

We are left with some freedom in the choice of $K_0$. Here we will take $K_0$ to be a function of $zp/M$ and $p/\mu$ for some new intermediate scale $\mu$.
The transformation in (\ref{symmetry}) then remains a symmetry of the action and we can repeat the previous ERG construction. To summarize, we consider
\begin{eqnarray}\label{dependence}
    K_s(zp/M),\qquad v_{ph}(p/\mu, zp/M)\ .
\end{eqnarray}
At this stage, there are two scales in the problem: 1) the renormalization scale $M/z$, and 2) an intermediate scale $\mu$ where we might allow the dispersion to transition, for example, from a non-relativistic to a relativistic form; see figure \ref{fig1}.

Before proceeding, let us note that this choice of regulator is natural if we consider the path integral in phase space or in first order terms. Then, 
\begin{equation}\label{pathintegralexcitedphasesp} 
\braket{\varphi(\vec x)|\Omega[b]}\sim\int [\mathcal{D}\pi\mathcal{D}\phi]_{\phi(-i\infty,\vec{x})\to 0}^{\phi(0,\vec{x})=\varphi(\vec{x})}\:\: e^{i (S_0+S_B+\mathcal{U}_b)}
\end{equation}
where 
\begin{equation}\label{phasespaction}
S^{nr}_0[\phi,\pi]=\int dt\Big( \pi\cdot \dot\phi - \tfrac12 \pi\cdot K_0\cdot \pi+\tfrac12\phi\cdot K_s^{-1}\vec\nabla^2\cdot \phi\Big).
\end{equation}
So we see that in phase space, $K_s$ and $K_0$ appear as separate regulator functions for the momentum modes of the field and its conjugate momentum field, respectively.

It is well-known that the ground state wavefunctional in a free theory is Gaussian. It can be written in the form
\begin{eqnarray}\label{gswf1}
     &&\braket{\varphi|\Omega}\sim\exp\lb -\frac{1}{2}\int_{\vec{p}} G^{-1} \varphi^a(\vec{p})\varphi^a(-\vec{p}) \rb, \nn\\
     &&G=z^{d-2}\frac{\sqrt{K_0K_s}}{p}=z^{d-2}K_s\frac{v_{ph}}{p}.
\end{eqnarray}

where in the second line we have written the kernel $G$ in terms of the functions $K_0,K_s$ defining the  path integral. In appendix \ref{app:canonical}, 
we verify this is the correct wave functional by canonically quantizing the Hamiltonian at a scale $z$.

The key observation is that the ground state wave functional is controlled by the combination $\tilde{K}=K_s v_{ph}$. To understand how this controls the entanglement of the state, it is conceptually useful to keep in mind the three distinct regions of momentum space defined by our set of regulators. Explicitly, these three regions are:
\begin{itemize}
    \item I: Low momentum modes with $p \leq \mu$,
    \item II: Intermediate momenta with $\mu \leq p \leq M/z$,
    \item III: High momentum modes with $p \geq M/z$.
\end{itemize}
As a regulator, $K_s$ effectively ``removes'' the high momentum modes in region III. Thus, the behavior of the ground state is essentially controlled by the regions I and II. In these regions we have $K_s \sim 1$, and so the kernel $G$ in the wavefunctional is entirely determined by the phase velocity $v_{ph}/p$. Note that in the UV, when $z$ is very small, region II's contribution to the wavefunctional dominates that of region I, whereas in the IR only region I contributes. With this in mind, we regard region II as controlling the UV behavior of the ground state while region I controls the IR behavior.

If we want the UV state to be unentangled in position space,  as in (\ref{unentangle}), we need the dispersion in region II to be such that $v_{ph}\sim p$ for almost all $p$.  In other words, the UV dispersion should be the non-relativistic $\omega(p)\sim p^2$, whereas a relativistic IR requires $\lim_{z\to \infty}v_{ph}=1$. These are requirements that are not hard to satisfy.
As an explicit example, consider the choice
\begin{eqnarray}\label{cMERAvph}
    v_{ph}(p/\mu, zp/M)^2=1+\lb \frac{M^2}{p^2z^2}\rb\lb \frac{p^4}{\mu^4}\rb\ .
\end{eqnarray}
Indeed, in the regime $\mu \ll p \ll M/z$, we have $\omega\sim \frac{M}{z\mu^2}p^2$ while $\omega\to p$ for $p\ll\mu$. So for $z\sim 1$, there is a large range of momentum over which the dispersion is non-relativistic, whereas as when $z$ has increased beyond $M/\mu$, a transition to a relativistic dispersion occurs. 
In appendix \ref{app:magic}, 
we establish that this choice corresponds to a particular construction discussed in \cite{zou2019magic} called {\it magic cMERA}.
A sketch can be found in Fig. \ref{fig1}.

What remains to be done is to repeat the ERG analysis for the generator of quantum states in the presence of the non-relativistic regulator. In fact, the analysis is a straightforward generalization, and we obtain (see appendix \ref{app:varyM})
\begin{eqnarray}\label{nonrelRG}
    z\p_z\braket{\varphi|\Omega[b]}=\bra{\varphi}\lb -\text{tr}_\mM\lb \tilde{\beta}[b]\cdot \frac{\delta}{\delta b}\rb+i \hat{\mathbf{K}} +i \hat{\mathbf{L}}\rb\ket{\Omega[b]}\nn\\
\end{eqnarray}
where
\begin{equation}
  \tilde\beta[b]=b\cdot \tilde\Delta_b\cdot b,
\qquad \tilde\Delta_b=M\p_M (\tilde D^{-2}).
\end{equation}
and
\begin{eqnarray}\label{newgzp}
    \tilde{g}_z=M\p_M \log (K_sv_{ph})\ .
\end{eqnarray}
We see that each of these kernels depends on both $K_0$ and $K_s$, or equivalently $K_s$ and $v_{ph}$. Specializing the analysis to the ground state by setting $b$ to zero, one indeed finds the ground state wave functional is of the form \eqref{gswf1}.

In comparing these non-relativistic flows to cMERA, it is clear that with choices such as the one in (\ref{cMERAvph}) the UV states are unentangled. What remains to be checked the locality of the disentangler.
For any non-relativistic flow, the expression for the kernel of the disentangler $g_z$ is given by the logarithmic derivative of $\tilde{K}=K_s v_{ph}$. It is this quantity which then decides the local properties of the coarse-graining. Given a choice such as that in \ref{cMERAvph}, the growth of $v_{ph}$ has the effect of widening the regulator in momentum space. This in turn sharpens the disentangler in position space, giving the desired real space coarse graining. On more general grounds, since $v_{ph}$ grows with $p$ to obtain a disentangled UV state, the cut-off function will always have some amount of smoothing, leading to a local kernel for the disentangler in position space. For some explicit examples, see appendix \ref{app:smoothgz}.

\section{Summary and discussions}\label{sec:conclude}

In summary, we have generalized the ERG to include non-relativistic flows of quantum states. In doing so, we have demonstrated by construction that cMERA can be defined intrinsically in the continuum from a microscopic description. By regulating the time and space derivatives in the Lagrangian separately, we have  allowed for a cut-off dependent dispersion relation that changes the real space entanglement pattern of the ground state. By choosing these two regulators such that the theory is spatially unentangled in the UV, we have recovered the usual coarse-graining operators found in earlier cMERA constructions. 

Perhaps the most promising feature of our generalized ERG is that it provides a systematic method to include interactions. Incorporating interactions in cMERA models using operator methods has proven to be difficult. For progress in this direction, see \cite{cotler2019entanglement,cotler2019renormalization,fernandez2019entanglement,fernandez2020non}. 
On the other hand, the path integral makes the incorporation of interactions straightfoward, at least in principle. The major technical difficulty is in finding appropriate limits where the ERG flow equations become tractable. While in this paper we explored the free fixed point, the ERG is also well behaved in large $N$ expansions and for certain exactly solvable deformations such as $T\bar{T}$ \cite{cavaglia2016tt}. These features of the ERG offer an exciting potential avenue to consistently define interacting cMERAs and ultimately make contact with holography, and the establishment of our result here is an important first step in that process.

Another interesting observation concerning our generalized RG flows is that they can accommodate more exotic unitary networks than cMERA. Indeed, by appropriately choosing the smooth cut-off functions $K_0$ and $K_s$, we can tune the properties of the ground state wavefunctional both in the UV and the IR. For example, we could construct a ground state which takes the form
\begin{equation}
    \langle\phi(\vec{x})\vert 0\rangle \propto \exp\lb -\frac{1}{2}\int_{\vec{p}} p^\alpha \varphi(-\vec{p})\varphi(\vec{p})\rb
\end{equation}
for any non-negative real number $\alpha$. The choice $\alpha = 1$ coincides with the relativistic free boson while $\alpha = 0$ yields the unentangled ground state of the non-relativistic free boson. Other choices correspond to Lifshitz-like dispersions $\omega\sim p^{1+\alpha}$. We leave the exploration of such networks, their entanglement properties, whether the operator algebra is of type I or type III, and the connection to recent work in \cite{Rottoli:2022plr} to future investigations. Note that the ERG formalism discussed here can accommodate any Gaussian theory, including theories without a local action such as generalized free fields (GFF).

It would also be interesting to further explore the connection between renormalization and error correction as put forth in \cite{Moosa2022}. There, the RG flows are viewed as an encoding channel for a particular set of low-energy states. How does the error correction picture generalize to the full ERG scheme, and what is the role of the large $N$ limit? By exploring large $N$ in both the interacting and non-interacting cases, we may use these continuous unitary networks as a toy model for better understanding the role of large $N$ in holographic systems in contrast to other features of holographic models such as large gap and strong coupling. 

Finally, in the ERG there are natural generalizations that lead to non-unitary flows which go beyond standard cMERA networks. In general, the expectation is that such ERG schemes correspond to a one-parameter family of quantum channels. These non-unitary channels may elucidate features of RG which connect to non-unitary entanglement renormalization schemes such as continuous Tensor Network Renormalization (cTNR), and the non-unitary generalization of MERA such as those discussed \cite{nozaki2012holographic,milsted2018geometric}. Furthermore, extending our formalism to non-unitary ERG schemes may shed light on questions regarding  RG monotones and irreversibility.

\begin{acknowledgments}
We thank Sumit Das, Keiichiro Furuya, Dominik Neuenfeld, Shoy Ouseph, Alfred Shapere, Matthew Thibodeau and Akash Vijay for discussions. Special thanks to Marc Klinger for notes on a second version of this manuscript. NL is grateful to the DOE for supporting this work through the grant DE- SC0007884 and the QuantiSED Fermilab consortium. The work of RGL was supported by the U.S. Department of Energy under contract DE-SC0015655. 
\end{acknowledgments}

\appendix

\onecolumngrid

\section{ERG for Non-Relativistic States}\label{app:varyM}

Let us recall the definition of the excited states we consider in the body of the letter:
\begin{eqnarray}
   && \braket{\varphi(\vec{x})\vert \Omega[b]}= \frac{1}{\sqrt{Z[b, \mathcal{U}_b]}}\int [\mathcal{D}\phi]^{\varphi} \exp\left(-S_{0}^{nr}[\phi, \varphi]- S_{b}\right),\nn\\
    &&S_{0}^{nr}=\frac{-1}{2z^{d-2}}\lb\int_{\mathcal{M}_-} \phi^a\cdot  \tilde{D}^2\cdot \phi^a -\int_\Sigma \varphi^a \cdot K_0^{-1} \partial_t\phi^a\Big|_{\Sigma}\rb,\nn\\
    &&S_{b}=\frac{1}{2z^{d-2}}\int_{\mathcal{M}_-}\phi^a\cdot b\cdot  \phi^a + \mathcal{U}_b\ .
\end{eqnarray}
By requiring these states to be normalized, the prefactor in the state is determined to be
\begin{equation}
    Z[b, \mathcal{U}_b] = \int [\mathcal{D}\phi] \exp\left(-\frac{1}{2z^{d-2}}\int_{\mathcal{M}} \phi^a\cdot( -\tilde{D}^2  + b)\cdot \phi^a - \mathcal{U}_b \right)
\end{equation}
This object may be regarded as the Euclidean ``partition function'' corresponding to the state $|\Omega[b]\rangle$. It is then natural to require that $Z[b, \mathcal{U}_b]$ satisfy the usual ERG condition $z\frac{d}{dz}Z[b, \mathcal{U}_b] = 0$. This will generally only be possible if the sources $b$ and $\mathcal{U}_b$ themselves flow as running couplings. We must be careful to do this in  a way that is consistent with the factorization of the path integral into states. This is done by first computing $z\frac{d}{dz}\langle \varphi(\vec{x}) \vert \tilde{b}\rangle$ directly, where $\ket{\tilde{b}} = \sqrt{Z[b, \mathcal{U}_b]} \ket{\Omega[b]}$ is the unnormalized path integral state. Then,
\begin{eqnarray}
    z\frac{d}{dz}Z[b, \mathcal{U}_b] &=& \int [\mathcal{D}\varphi]\left(z\frac{d}{dz}\langle \tilde{b} \vert \varphi(\vec{x})\rangle \right)\langle \varphi(\vec{x}) \vert\tilde{b}\rangle 
    +\int [\mathcal{D}\varphi]\:\langle \tilde{b} \vert \varphi(\vec{x})\rangle z\frac{d}{dz}\langle \varphi(\vec{x}) \vert \tilde{b}\rangle =0 .
\end{eqnarray}
Taking the derivative of the unnormalized path integral state yields
\begin{equation}
    z\frac{d}{dz}\langle \varphi(\vec{x})\vert \tilde{b}\rangle  = -\int [\mathcal{D}\phi]^\varphi \left(z\frac{d}{dz}S_0^{nr} + z\frac{d}{dz}S_{b}\right)e^{-S_0^{nr} - S_{b}}
\end{equation}
Evaluating the derivatives explicitly, one finds:
\begin{eqnarray}\label{S-source-derivative}
    z\frac{d}{dz}S_{b} &=& \frac{1}{2z^{d-2}}\int_{\mathcal{M}_-} \phi^a\cdot \left(z\frac{d}{dz}b\right)\cdot \phi^a - \frac{d-2}{2z^{d-1}}\int_{\mathcal{M}_-} \phi^a \cdot b\cdot \phi^a +z\frac{d}{dz}\mathcal{U}_b,\\  \label{S0-derivative}
  z\frac{d}{dz}S_0^{nr}
    &=& -\frac{1}{2z^{d-2}}\int_{\mathcal{M}_-}\phi^a\cdot \tilde{D}^2\cdot \Delta_b\cdot \tilde{D}^2\cdot \phi^a-  \frac{1}{2z^{d-2}}\int_{\Sigma}\varphi^a\cdot g\cdot K_0^{-1}\left(\frac{z^2 \nabla^2}{M^2}\right)\partial_t\phi^a\Big\vert_\Sigma\\&&\hspace{2cm}-\frac{(d-2)}{2z^{d-1}}\int_{\mathcal{M}_-}\phi^a\cdot \tilde{D}^2\cdot \phi^a-\frac{(d-2)}{2z^{d-1}}\int_\Sigma \varphi^a\cdot K_0^{-1} \cdot \partial_t \phi^a\Big\vert_\Sigma
\end{eqnarray}
where we have defined the bilocal kernels
\begin{align}
    \Delta_b(x, y) = z\frac{d}{dz}(\tilde{D}^2)^{-1}(x, y), \hspace{2cm}
    g(\vec{x}, \vec{y}) = \left(z\frac{d}{dz}\log\left(K_0\left(z^2\nabla^2/M^2\right)\partial_t\right)\right)(\vec{x}, \vec{y}).
\end{align}
To proceed we will need to use a result which is straightforward to prove (see for example Appendix D.1 of \cite{fliss2017unitary}). For our path integral states, the identity is a Schwinger-Dyson equation for operator insertions $\mathcal{O}$ in the path integral which reads
\begin{eqnarray}
    \mathcal{O}\frac{\delta S_0^{nr}}{\delta \phi} \sim \frac{\delta \mathcal{O}}{\delta \phi} - \frac{\delta S_{b}}{\delta \phi}.
\end{eqnarray}
Applying this identity twice to the first term of (\ref{S0-derivative}), one obtains
\begin{eqnarray}
    &&-\frac{1}{2z^{d-2}}\int_{\mathcal{M}_-} \phi^a \cdot \tilde{D}^2 \cdot \Delta_b\cdot \tilde{D}^2 \cdot \phi^a \sim  -\frac{1}{2z^{d-2}}\int_{\mathcal{M}_-}[\tr(\tilde{D}^2\cdot \Delta_b) - \tr(\Delta_b\cdot b) + \phi^a\cdot b\cdot \Delta_b\cdot b\cdot \phi^a]
\end{eqnarray}

Similarly in the third term of (\ref{S0-derivative}), we have
\begin{eqnarray}
    &&-\frac{d-2}{2z^{d-1}}\int_{\mathcal{M}_-}\phi^a\cdot \tilde{D}^2 \cdot \phi^a= -\frac{d-2}{2z^{d-1}}\left(\tr(\text{Id}_{\mathcal{M}_-}) - \int_{\mathcal{M}_-} \phi^a \cdot b\cdot \phi^a\right)
\end{eqnarray}
where $\text{Id}_{\mathcal{M}_-}$ is the delta function restricted to the lower half plane. 
By comparing all the terms we have written out so far, we see that all the quadratic insertions in the integrand are canceled if the source $b$ satisfies
\begin{equation}
    z\frac{d}{dz}b =  \int_{\mathcal{M}_-}b\cdot \Delta_b\cdot b
\end{equation}
Since quadratic terms contain no boundary dependence and using the fact that $Z[b, \mathcal{U}_b]$ is reflection symmetric, this result is also the necessary flow equation to ensure such terms cancel in the expression for the partition function as well. What remains then are the normalizations and a boundary term.
\begin{eqnarray}
    &&z\frac{d}{dz}\langle \varphi(\vec{x})\vert \tilde{b}\rangle = -\int [\mathcal{D}\phi]^\varphi\: e^{-S_0^{nr} - S_{b}}\Bigg(z\frac{d}{dz}\mathcal{U}_b + \frac{1}{2}\int_\Sigma \varphi^a\cdot \Delta_\Sigma\cdot \frac{\delta}{\delta \varphi^a}- \frac{1}{2z^{d-2}}\int_{\mathcal{M}_-} \tr\left(\tilde{D}^2\cdot \Delta_b - \Delta_b  \cdot b+ \frac{d-2}{z}\text{Id}_{\mathcal{M}_-}\right)\Bigg)\nn\\
\end{eqnarray}
where we have also defined the boundary kernel $\Delta_\Sigma(\vec{x}, \vec{y}) = g(\vec{x}, \vec{y}) + \frac{d-2}{z}\delta^{(d-1)}(\vec{x}-\vec{y})$. 
Plugging this into the expression for $Z[b, \mathcal{U}_b]$, we then obtain
\begin{eqnarray}
    &&z\frac{d}{dz}Z[b, \mathcal{U}_b]= -\int[\mathcal{D}\varphi][\mathcal{D}\phi] \: e^{-{S_0^{nr}}^\dagger - S_{b}^\dagger} \Bigg(2z\frac{d}{dz}\mathcal{U}_b - \frac{1}{z^{d-2}}\int_{\mathcal{M}_-} \tr\left(\tilde{D}^2 \cdot \Delta_b - \Delta_b\cdot b + \frac{d-2}{z}\text{Id}_{\mathcal{M}_-}\right)\nn\\
    &&\hspace{7cm}+\frac{1}{2}\int_\Sigma\Big(\varphi^a\cdot \Delta_\Sigma  \cdot \frac{\delta}{\delta \varphi^a} + \stackrel{\leftarrow}{\frac{\delta}{\delta \varphi^a}}\cdot \
    \Delta_\Sigma\cdot \varphi^a\Big)\Bigg)e^{-S_0^{nr} - S_{b}}\nn\\
\end{eqnarray}
The dagger in the above expression denotes Euclidean time reflection. The difference in order in the last two terms ensures that the time derivative pulled down comes from the correct half of the Euclidean plane. But, we may then integrate by parts in the path integral, producing the commutator $\left[\varphi^a(\vec{x}), \frac{\delta}{\delta \varphi^a(\vec{y})}\right] = -\delta^{(d-1)}(\vec{x}, \vec{y})$. We then conclude that invariance of the partition function demands
\begin{eqnarray}
    z\frac{d}{dz}\mathcal{U}_b &=& \frac{1}{2z^{d-2}}\int_{\mathcal{M}_-}\tr(\tilde{D}^2\cdot \Delta_b - \Delta_b \cdot b + \frac{d-2}{z}\text{Id}_{\mathcal{M}_-}) + \frac{1}{4}\int_\Sigma \tr(\Delta_\Sigma)\ .
\end{eqnarray}
Using this last relation, we may finally conclude that the normalized wavefunctional satisfies the flow equation
\begin{eqnarray}
    z\frac{d}{dz}\langle \varphi(\vec{x})\vert  \Omega[b](z)\rangle &=&  \int [\mathcal{D}\phi]^\varphi \:e^{-S_0^{nr} - S_{b}}\left(\frac{1}{2}\int_\Sigma \varphi^a \cdot g\cdot \frac{\delta}{\delta \varphi^a} + \frac{1}{4}\int_\Sigma \tr(g)\right)\nn\\
\end{eqnarray}
or, written more symmetrically,
\begin{eqnarray}
    z\frac{d}{dz}\langle \varphi(\vec{x})\vert  \Omega[b](z)\rangle &=  &\int [\mathcal{D}\phi]^\varphi\:e^{-S_0^{nr} - S_{b}}\left(\frac{1}{4}\int_\Sigma \left[\varphi^a \cdot g \cdot \frac{\delta}{\delta \varphi^a} + \frac{\delta}{\delta \varphi^a}\cdot g \cdot \varphi^a\right]\right)\nn\ .\\
\end{eqnarray}

\section{Quasi-local disentanglers}\label{app:smoothgz}

The equation (\ref{gdelta}) relates the kernel of the disentangler in (\ref{cond3}) to the smooth regulator function $K$:
\begin{eqnarray}\label{gzpapp}
    g_z=M\p_M \log K\ .
\end{eqnarray}
It is natural to choose functions $K$ which smoothly approximate a Heaviside function centered at the effective scale $M/z$. In this Appendix, we explore two such choices of $K$ that lead to quasi-local disentanglers. The first choice of $K$ suppresses the high momentum modes exponentially, and the other choice suppresses them by a power law:
 \begin{eqnarray}\label{Kchoices}
     &&K_{exp,\alpha}(zp/M)=\frac{1+e^{-\alpha}}{1+e^{\alpha(z p/M -1)}}\nn\\
     &&K_{pl,\alpha}(zp/M)=\frac{1}{1+(z p/M)^{2\alpha}},
 \end{eqnarray}
 and both tend to a sharp cut-off in the limit $\alpha\to \infty$. 

\begin{figure}
\centering
     \includegraphics[width=.6\linewidth]{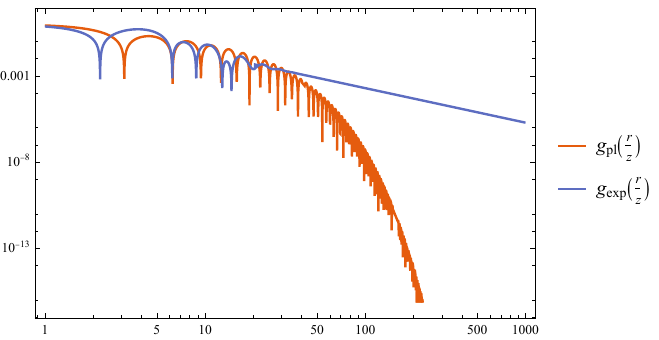}
\caption{Log-log plot of the position space kernels $g_{pl}$ in orange and $g_{exp}$ in blue. Power law suppression in momentum space leads to exponential fall off in position space and vice versa. Plot values are $\alpha = 10$ and $d =1$. }
\label{fig2}
\end{figure}

 It follows from (\ref{gzpapp}) that these choices correspond to the disentanglers with the kernel 
 \begin{eqnarray}
     &&g_{exp,\alpha}(p)=\frac{\alpha z p}{M}\lb 1-\frac{1}{1+e^{\alpha(z p/M-1)}}\rb\nn\\
     &&g_{pl,\alpha}(p)=2\alpha\lb 1-\frac{1}{1+(z p/M)^{2\alpha}}\rb\ ,
 \end{eqnarray}
 respectively.
 In real space they become
 \begin{eqnarray}
     &&g_{exp,\alpha}(\vec{r})=\frac{\alpha z}{M|r|}\lb \delta(\vec{r})-\frac{1}{1+e^{\alpha(z|\nabla_r|/M-1)}}\rb\nn\\
     &&g_{pl,\alpha}(\vec{r})=2\alpha \lb \delta(\vec{r})-\frac{1}{1+(z^2\nabla^2_r/M^2)^\alpha}\rb
 \end{eqnarray}
 and their disentanglers are
 \begin{eqnarray}
     \hat{\mathbf{K}}(z) &= \frac{1}{4}\int \lb\hat{\varphi}^a(\vec{x})g_z(\vec{r}) \hat{\pi}^a(\vec{x}+\vec{r})+h.c.\rb\ .
 \end{eqnarray}
 The term proportional to the $\delta(\vec{r})$ is manifestly quasi-local, so we
 only need to consider the remaining non-local term. In position space, the exponential and the power law cases, respectively, give
 \begin{eqnarray}
  && \frac{\alpha z}{(2\pi M)^{d/2}r^{d/2-1}}\int_0^\infty dp\: \frac{p^{d/2+1}}{1+e^{\alpha(pz/M-1)}} J_{d/2-1}(p r/M)\nn\\
   &&\frac{2\alpha}{(2\pi)^{d/2}r^{d/2-1}}\int_0^\infty dp\: \frac{p^{d/2}}{1+(z p/M)^{2\alpha}} J_{d/2-1}(pr/M) \ .
\end{eqnarray}
Both integrals are oscillating functions of $r/z$ with an amplitude that decays. As we increase $\alpha$ and $z$ the disentangler kernel $g_z(\vec{r})$ becomes more non-local. However, for all values of $\alpha$,  the decay of $g_{exp}$ in $r$ is power-law, whereas  the decay of $g_{pl}$ is exponential in $r$; see Fig. \ref{fig2}.

\section{Calculating the ground state wave functional}\label{app:transamapp}

Here, we calculate the ground state wave functional in the theory with the non-relativistic cut-offs $K_0$ and $K_s$ at any $z$ by evaluating the path integral for generic transition amplitudes first and then taking a limit. This discussion closely parallels the one presented in \cite{fliss2017unitary}.

Consider the transition amplitude between times $t_-$ and $t_+$ with fixed field configurations $\varphi_-$ and $\varphi_+$, respectively:

\begin{equation}
\Big\langle\varphi_+,t_+\Big|\varphi_-,t_-\Big\rangle\equiv \mathcal Z[\varphi_\pm]\equiv\mathcal N\int \left[\mathcal{D}\phi\right]^{\varphi_+}_{\varphi_-}e^{iS_0^{nr}}
\end{equation}
with the non-relativistically regulated action
\begin{eqnarray}
S_0^{nr}&=&\frac{1}{2z^{d-2}}\int_{t_-}^{t_+}dt\int d^{D}\vec x\;\phi(t,\vec x)\left(-K_0^{-1}\left(-\frac{z^2}{M^2}\vec D^2\right)D_t^2+K_s^{-1}\left(-\frac{z^2}{M^2}\vec D^2\right)\vec D^2\right)\phi(t,\vec x)
\\&&\qquad+\sum_\pm\pm\frac{1}{2z^{d-2}}\int_\Sigma d^{D}\vec x\;\varphi_{\pm}\cdot K_0^{-1}\left(-\frac{z^2}{M^2}\vec D^2\right)\cdot\left.D_t\phi\right|_{t_\pm}\ .
\end{eqnarray}

The normalization of the path integral $\mathcal N$ is fixed using the conditions of orthonormality and factorization by a complete set of states \cite{fliss2017unitary}.

We make the field redefinition $\phi=\phi_c+\chi$ with $\phi_c$ the classical solution to the equations of motion subject to the boundary conditions $\varphi_\pm$ and $\chi$ the quantum fluctuations forced to zero at the boundary.  The classical solution is
\begin{eqnarray}
    \phi(t,\vec x)=\int\frac{d^Dp}{(2\pi)^D}e^{i\vec p\cdot\vec x}\left[e^{i\omega_{\vec p}(t-t_-)}\frac{e^{i\omega_{\vec p}T}\varphi_+(\vec p)-\varphi_-(\vec p)}{2ie^{i\omega_{\vec p}T}\sin\omega_{\vec p}T}
-e^{-i\omega_{\vec p}(t-t_-)}\frac{e^{-i\omega_{\vec p}T}\varphi_+(\vec p)-\varphi_-(\vec p)}{2ie^{-i\omega_{\vec p}T}\sin\omega_{\vec p}T}\right]_{\omega_{\vec p}=\sqrt{\frac{K_0}{K_s}}p}
\end{eqnarray}
which gives
\begin{eqnarray}
\dot\phi(t_+,\vec p)=\frac{\omega_{\vec p}}{\sin\omega_{\vec p}T}\left[\varphi_+(\vec p)\cos\omega_{\vec p}T-\varphi_-(\vec p)\right],\qquad
\dot\phi(t_-,\vec p)=\frac{\omega_{\vec p}}{\sin\omega_{\vec p}T}\left[\varphi_+(\vec p)-\varphi_-(\vec p)\cos\omega_{\vec p}T\right]
\end{eqnarray}
The shift of the field leaves a path integral over $\chi$ with Dirichlet boundary conditions times a classical boundary action
\begin{align}
S_B=&\sum_\pm\pm\frac{1}{2z^{d-2}}\int_\Sigma d^{D}\vec x\;\varphi_\pm K_0^{-1}\left.D_t\phi_c\right|_{t_\pm}\nonumber\\
=&\frac{1}{2z^{d-2}}\int \frac{d^{D}\vec p}{(2\pi)^{D}}\frac{\omega_{\vec p}}{\sin(\omega_{\vec p}T)}K_0^{-1}\left(\frac{z^2}{M^2}{\vec p}^2\right)\left(\cos(\omega_{\vec p}T)\left(\varphi_+(\vec p)\varphi_+(-\vec p)+\varphi_-(\vec p)\varphi_-(-\vec p)\right)-2\varphi_+(\vec p)\varphi_-(-\vec p)\right).
\end{align}
where $T=t_+-t_-$ and 
\begin{eqnarray}
\omega_{\vec p}^2=\frac{K_0}{K_s} p^2.    
\end{eqnarray}
 The path integral over $\chi$ can be evaluated using eigenfunctions of $\vec D^2$ giving the explicit expression for the free transition amplitude (see \cite{fliss2017unitary} for more detail)  

\begin{align}\label{explicitfreeamp}
\mathcal Z[\varphi_+,\varphi_-]=&\prod_{\vec p}\left(\frac{z^{2-d}K_0^{-1}\omega_{\vec p}}{i2\pi\sin(\omega_{\vec p}T)}\right)^{N/2}\nonumber\\
&\times\exp\left(\frac{i}{2z^{d-2}}\int \frac{d^{D}\vec p}{(2\pi)^{D}}\frac{\omega_{\vec p}}{\sin(\omega_{\vec p}T)}K_0^{-1}\left(\cos(\omega_{\vec p}T)\left(\varphi_+(\vec p)\varphi_+(-\vec p)+\varphi_-(\vec p)\varphi_-(-\vec p)\right)-2\varphi_+(\vec p)\varphi_-(-\vec p)\right)\right)
\end{align}
which is Gaussian in $\varphi_\pm$.

Consider the case where the contour is purely Euclidean, running from $t_i=iT$ to $t_f=-iT$. As discussed in \cite{fliss2017unitary}, in the limit $T\rightarrow\infty$ the fall off of the transition amplitude with $T$ factors into pieces given by the ground state wave-functinoal.  Using \eqref{explicitfreeamp} we have
\begin{align}
\lim_{\beta\rightarrow\infty}\mathcal Z[\varphi_+,\varphi_-]=&\prod_{\vec p}\left(\frac{z^{2-d}K_0^{-1}\omega_{\vec p}}{\pi e^{2\omega_{\vec p}T}}\right)^{N/2}\!\!\!\!\!\!\!\!\times\exp\left(-\frac{1}{2z^{d-2}}\!\!\int\!\!\frac{d^{D}\vec p}{(2\pi)^{D}}K_0^{-1}\left(\frac{z^2}{M^2}{\vec p}^2\right)\omega_{\vec p}\left(\varphi_+(\vec p)\varphi_+(-\vec p)+\varphi_-(\vec p)\varphi_-(-\vec p)\right)\right)\nonumber\\
=&e^{-2T\,E_\Omega}\Psi^*_\Omega[\varphi_+]\Psi_\Omega[\varphi_-].
\end{align}
From here it is easy to isolate the expression for the ground state wave functional in the regulated theory:
\begin{equation}\label{appGSWF}
\Psi_\Omega[\varphi]=\prod_{\vec p}\left(z^{2-d}\pi^{-1}\omega_{\vec p}\,K_0^{-1}\right)^{N/4}\times\exp\left(-\frac{1}{2z^{d-2}}\int\frac{d^{D}\vec p}{(2\pi)^{D}}\sqrt{\frac{\vec p^2}{K_0K_s}}\,\varphi(\vec p)\varphi(-\vec p)\right).
\end{equation}

\section{Canonical Quantization and Operator Renormalization}\label{app:canonical}

A complementary viewpoint is obtained by considering the canonical quantization of the model \eqref{actionregnew} at a given scale set by $z$. This can be interpreted as a Gaussian deformation 
\begin{eqnarray}\label{deform}
S_{def}=\int_x \pi^a(x)K_s^{-1}(v_{ph}^{-2}-1)\pi^a(x) 
\end{eqnarray}
away from the relativistically dispersing theory 
that fine-tunes the dispersion $v_{ph} p$ to the desired form.
As in \eqref{phasespaction}, the Hamiltonian corresponding to the regulated action is quadratic,
\begin{eqnarray}\label{Hamil}
     H(z)&=&\frac{1}{2}\int _{\vec{p}}\lb K_0\pi^a(-\vec{p})\pi^a(\vec{p}) +K_s^{-1}p^2\varphi^a(-\vec{p})\varphi^a(\vec{p}) \rb\nn\\
     &=&\frac{1}{2}\int _{\vec{p}}p \: v_{ph}\lb G\pi^a(-\vec{p})\pi^a(\vec{p}) +G^{-1} \varphi^a(-\vec{p})\varphi^a(\vec{p}) \rb
\end{eqnarray}
where $G=z^{d-2}\tilde{K}/p$ is the same distribution appearing in (\ref{gswf1}). 
The factors of $z^{d-2}$ in $G$ correspond to the canonical dimension of the field under a Weyl rescaling. It will lead to the scaling operator $\hat{\mathbf{L}}$ in the RG flow. 

We canonically quantize the theory by defining the scale-dependent annihilation operator
\begin{eqnarray}\label{Hamiltoncanon}
     &&\hat b^a_z(\vec{p})=\frac{1}{\sqrt{2}}\Big(G^{-1/2}\:\hat \varphi^a(\vec{p})+i G^{+1/2}\hat \pi^a(\vec{p})\Big)\nn\\
     &&\hat H(z)=\frac{1}{2} \int_{\vec{p}}\omega_{\vec p}\: \hat b^a_z(\vec{p})^\dagger \hat b_{a,z}(\vec{p})+E_\Omega\hat{I}
\end{eqnarray}
which depends on $K_0$ and $K_s$ only through the combination $G$. We can think of the act of increasing $z$ as corresponding \cite{Minic:1994ff,Alexanian:1998wu} to a canonical transformation $\hat U_z$ that renormalizes the free field and its canonical conjugate momentum field according to
\begin{eqnarray}
     &&\hat b_z=\hat U_z\hat  a\hat U_z^\dagger\nn\\
       &&\hat U_z \hat \varphi^a(\vec{p}) \hat U_z^\dagger =G^{-1/2} \hat \varphi^a(\vec{p})\nn\\
     &&\hat U_z \hat \pi^a(\vec{p})\hat U_z^\dagger=G^{1/2}\hat \pi^a(\vec{p})\ .
\end{eqnarray}
Such unitary transformations are called the unitary squeezing operators:
\begin{eqnarray}\label{defRG}
   &&\hat U_z=\prod_{\vec{p}}\hat S_{\vec{p}}\lb-\tfrac{1}{2}\log G\rb\nn\\
     &&\hat S_{\vec{p}}(\xi):=\exp\lb\xi \hat a_a^\dagger(\vec{p})\hat a_a^\dagger(-\vec{p})-\xi^* \hat a_a(\vec{p})\hat a_a(-\vec{p})\rb\nn\\
     &&\hat a(\vec{p})=\sqrt{\frac{p}{2}}\hat \varphi_a(\vec{p})+\frac{i\hat \pi_a(\vec{p})}{\sqrt{2p}}\ .
\end{eqnarray}
More explicitly,
\begin{eqnarray}
    \hat U_z=\exp\lb \frac{i}{2}\int_{\vec{p}} \log G\lb \hat \varphi(-\vec{p})\hat \pi(\vec{p})+\hat \pi(\vec{-p})\hat \varphi(\vec{p})\rb \rb\nn\ .
\end{eqnarray}
The vacuum at scale $z$ is related to the free field vacuum by $\ket{\Omega(z)}=\hat U_z\ket{\Omega}$ because it is annihilated by all $\hat b_z^\alpha(\vec{p})$. In other words, the vacuum at scale $z$ satisfies
\begin{eqnarray}
     \lb \hat \varphi^a(\vec{p})+iG\hat \pi^a(\vec{p})\rb\ket{\Omega(z)}=0
\end{eqnarray}
which is solved by the vacuum-squeezed wave functional in (\ref{gswf1}).

We can write down (\ref{defRG}) as a unitary flow 
\begin{eqnarray}\label{flowU}
    &&\hat U_z=e^{i\int^z_1 \frac{dz'}{z'} (\hat{\mathbf{K}} (z')+\hat{\mathbf{L}})}
\end{eqnarray}
where $\hat{\mathbf{L}}$ is the scaling operator in (\ref{condL}), and $\hat{\mathbf{K}}$ is the disentangler in (\ref{cond3}) with the new kernel in (\ref{newgzp}) in exact analogy with (\ref{nonrelRG}). Note that if the phase velocity $v_{ph}$ is chosen to be independent of $z$, one has to multiply \eqref{flowU} by an initial unitary that corresponds to an irrelevant deformation at $z=1$.

\section{Magic cMERA}\label{app:magic}

In \cite{zou2019magic} an example of cMERA for two-dimensional massless free bosons was constructed with the special property that at any point along the flow, the state $\hat U_z\ket{\Lambda}$ was the exact ground state of a strictly local Hamiltonian $H(z)$:
\begin{eqnarray}\label{magiccMERA}
  &&\hat  H(z)=\frac{1}{2}\int \lb \hat\pi(x)^2+(\p_x\hat\phi)^2+\frac{1}{z^2\Lambda^2}(\p_x\hat\pi)^2\rb.
\end{eqnarray}
This construction was called  magic cMERA.
It corresponds to an irrelevant deformation of the theory similar to (\ref{deform}) that  modifies the dispersion relation in the UV to allow for unentangled states.
To see this explicitly, we rewrite the Hamiltonian in momentum space, and notice that it takes the form in (\ref{Hamil}) for the choice
\begin{eqnarray}
    &&K_0=1+\frac{p^2}{z^2\Lambda^2}, \qquad K_s=1\ .
\end{eqnarray}
The dispersion relation is modified to
\begin{eqnarray}\label{vphasmagiccMERA}
    v_{ph}=\sqrt{1+\frac{p^2}{z^2\Lambda^2}}
\end{eqnarray}
which has the desired form both in the IR $\lim_{z\to \infty}v_{ph}\to 1$, and in the UV it becomes non-relativistic
\begin{eqnarray}
    z\ll 1:\qquad v_{ph}\sim \frac{p}{z\Lambda}\ .
\end{eqnarray}
Note that this deviates from our discussion of cMERA up to here, because we were requiring the state at $z=1$ to be the unentangled nonrelativistic ground state $\ket{\Lambda}$. To arrange for $\ket{\Lambda}$ at $z=1$, the authors of \cite{zou2019magic} also considered a deformation of (\ref{magiccMERA}) by a mass term $\Lambda \hat{\phi}^2$.

Deep in the UV the ground state wave functional in (\ref{gswf1}) factors  in real space as in (\ref{unentangle}) as desired. Comparing to our ERG construction, we make the following two observations. 
Note that  the dispersion of magic cMERA in (\ref{vphasmagiccMERA}) is the special case in (\ref{cMERAvph}) 
with $\Lambda=\mu^2/M$.
Second, in our non-relativistic ERG, $K_s$ played the role of a smooth regulator. Here, since $K_s=1$ the magic cMERA corresponds to an unregulated version of the theory. Thus, the ERG construction can be thought of as an elaboration in the context of a properly regulated theory.

\bibliographystyle{apsrev4-1}

\bibliography{letter}

\end{document}